\documentclass[final,3p,times,twocolumn]{elsarticle}
\usepackage{amssymb}
\usepackage{multicol}
\usepackage{graphicx}
\usepackage{epsfig}
\usepackage{fancybox}
\usepackage{pstricks,pst-node,pst-text,pst-3d}
\usepackage{subfigure}
\journal{Journal of Magnetism and Magnetic Materials}
\begin{document}
\begin{frontmatter}
\title{Time dependent magnetic field effects on the $\pm J$  Ising model}
\author{Erol Vatansever$^{a,b}$}
\author{Umit Akinci$^{a}$}
\author{Hamza Polat$^{a,\dagger}$}
\ead{hamza.polat@deu.edu.tr}
\cortext[cor1]{Corresponding Author. Tel:+90-232 4128672 Fax:+90-232 4534188}
\address[label1]{Department of Physics, Dokuz Eyl\"{u}l University, TR-35160 Izmir, Turkey}
\address[label2]{Dokuz Eyl\"{u}l University, Graduate School of Natural and Applied Sciences, TR-35160 Izmir, Turkey}
\begin{abstract}
Nonequilibrium phase transition properties of the $\pm J$ Ising model
under a time dependent oscillating perturbation
are investigated within the framework of  effective field
theory for a two-dimensional square lattice. After a detailed
analysis, it is found that  the studied system exhibits
unusual and interesting behaviors such as reentrant phenomena,
and the competition between  ferromagnetic and
antiferromagnetic exchange interactions gives rise to destruct
the dynamic first  order phase transitions as well as
dynamic tricritical points. Furthermore, according to N\'{e}el nomenclature,
the magnetization  profiles have been found
to obey Q-type, L-type  and P-type classification schemes
under certain conditions. Finally, it is  observed that the
treatment of critical percolation with  applied field
amplitude strongly depends upon the  frequency of time
varying external field.
\end{abstract}
\end{frontmatter}
\section{Introduction}
Investigation of disorder effect problems which may be originated
from random interactions between spins or from a random
dilution of the magnetic ions with  non-magnetic species on
the magnetic materials has a long history, and there have
been a great many studies  focused on disordered magnetic
materials with  quenched  randomness where the random variables
of a magnetic system  may not change its value over  time. Actually,
all magnetic materials have  some small  defects, and  magnetic
properties, i.e phase transition temperature point that
separates the ordered phase from disordered phase, of sample
varies  significantly  depending on  the type of defects. From
the experimental point of view, a number of studies including
the site or bond randomness have been devoted to better
understanding  of magnetic properties of different types of
magnetically interacting systems such as
$\mathrm{Rb_{2}Mn_{c}Mg_{1-c}F_{4}}$ \cite{birgeneau},
$\mathrm{Mn_{p}Zn_{1-p}F_{2}}$ \cite{baker},
$\mathrm{Fe_{p}Mg_{1-p}Cl_{2}}$ \cite{wood}, $\mathrm{Cd_{1-p}Mn_{p}Te}$
\cite{oseroff}, $\mathrm{Mn_{c}Zn_{1-c}F_{2}}$ \cite{cowley},
$\mathrm{Co(S_{p}Se_{1-p})_{2}}$ \cite{adachi}, and
$\mathrm{Co_{p}Zn_{1-p}(C_{5}H_{5}NO)_{6}(ClO_{4})_{2}}$ \cite{algra}.

From the theoretical point of view, a great deal of studies have
been performed regarding the magnetic properties of disordered
magnetic systems,  and based on the investigation of static or
dynamic phase transition (DPT) properties of such disordered magnetic
system, theoretical works can be classified in two basic categories.
In the former group,  static or equilibrium properties  of
these type of  systems have been analyzed within the
several frameworks such as series expansion method \cite{elliot,
domb, sykes, jakubczak},  Monte Carlo method \cite{frisch, vyssotsky,
kirkpatrick, stauffer, heuer},  renormalization  group theory \cite {reynolds, plischke,
yeomans, yeomans2, klein, barma, levy, tsallis}, replica method
\cite{giri, dominicis, domany}, effective field theory \cite{kaneyoshi,
almeida, kaneyoshi2, fittipaldi, taggart, bonfim, honmura1, sarmento,
balcerzak, sarmento2, bobak, zheng} and bethe method \cite{katsura}.

The statistical mechanics of nonequilibrium systems
is a less well developed and understood field of study  than that for
equilibrium systems. Even though the physical investigations
regarding these systems bring about a lot of mathematical difficulties,
the nonequilibrium systems are in the focus of scientists because
they have an exotic, unusual and interesting dynamic behaviors.
For instance, the universality class of the kinetic Ising model
under a time dependent  oscillating magnetic  field is different
from its equilibrium  counterparts \cite{chakrabarti, acharyya}.
It is exciting to say that a magnetic system composed of interacting
magnetic moments under the influence of a time dependent magnetic
field exhibits two important phenomena: DPT and
dynamic  hysteresis behavior. Dynamic evolution of the system
strongly depends upon the amplitude and frequency of the applied
external magnetic field as well as other intrinsic or extrinsic
Hamiltonian parameters. It is possible to mention that nonequilibrium
phase transitions originate due to a competition between time scales
of the relaxation time of the system and oscillating period of the
external field. As far as we know, the ferromagnetic system exists in
dynamically disordered phase at high temperatures and for high
amplitudes of the periodically varying magnetic field.  In this region,
the time dependent magnetization is able to follow the external magnetic
field with some delay whereas this is not the case for low temperatures
and small magnetic field amplitudes. The mechanism shortly described above
points out the existence a  DPT \cite{chakrabarti}, and
theoretical point of view, a number of studies concerning the DPTs
as well as hysteresis  properties of different type
systems under the time dependent perturbation  have been performed by using
well known methods such as Monte Carlo simulation \cite{lo, acharyya2,
acharyya3, acharyya4, rao, sides, zhu, park}, effective field
theory \cite{shi, deviren, yuksel}, mean field theory \cite{chakrabarti,
acharyya4, acharyya5, punya} as well as hard spin mean field theory
\cite{sariyer}.

In addition to the theoretical published works mentioned above,
the DPTs and also hysteresis behaviors
can be observed experimentally due to recent developments
in experimental techniques. It is beneficial  to give a few examples
regarding these properties such as epitaxial Fe/GaAs(001) and
Fe/InAs(001) ultrathin films \cite{moore}, finemet thin
films \cite{santi}, [Co(4$\mathrm{A^{o}}$)/Pt(7$\mathrm{A^{o}}$)]
multi-layer system with strong perpendicular anisotropy \cite{robb},
Co films on a Cu (001) surface \cite{jiang}, thin polycrystalline
Ni$_{80}$Fe$_{20}$ films \cite{choi}, Fe$_{0.42}$Zn$_{0.58}$F$_{2}$
\cite{rivera}, epitaxial Fe/GaAs(001) thin films \cite{lee},
epitaxial single ferromagnetic  fcc NiFe(001), fcc Co(001), a
nd fcc NiFe/Cu/Co(001) layers \cite{lee2} as well as ultrathin
ferromagnetic Fe/Au(001) films \cite{he}.  Based upon the
detailed  experimental investigations, it has been discovered
that  experimental nonequilibrium dynamics of considered real
magnetic systems strongly  resemble the dynamic behavior predicted
from theoretical calculations of a  kinetic  Ising model. From
this point of view,  it is possible to see  that  there exists an
impressive  evidence of qualitative  consistency  between theoretical
and experimental  investigations.

On the other hand, in the latter group there exists a
limited number of  nonequilibrium studies  concerning randomness
effects. For instance,  by making use of  both mean field theory and Monte Carlo
simulation method,   it has been shown that  the hysteresis loop
area of a ferromagnet including impurities under a time dependent
magnetic field is a power law function of the linear driving rate
as $A-A_{0}\propto h^{\beta}$, where, $A_{0}, h$ and  $\beta$ are
the  static hysteresis loop area, the linear driving rate and
scaling exponent of the  system, respectively \cite{zheng_li}. Very
recently,  DPT  and also hysteretic properties
such as remanence, coercivity and loop area of the random quenched
site kinetic Ising model have been probed based on the effective field
theory with single-site correlations and it is observed that the
coexistence region, where dynamically ordered and disordered phase
coexist depending on  selected Hamiltonian parameters, disappear for
sufficiently weak dilution of  lattice sites. It is also propounds
that the considered disordered system indicates  the existence
of essentially three,  particularly four types of dispersion
curves \cite{akinci, aktas}. Following the same methodology,
quenched random bond diluted Ising model, where the spin-spin exchange
interaction  has a probability $p$ and $1-p$ of taking on
values $J$ and $0$, respectively, has been analyzed by benefiting
from Glauber type stochastic model \cite{glauber}, and particular
attention has been devoted the better understanding of effects of
bond randomness on the global phase diagrams constructed in related
planes and on the microscopic origin of the  magnetic system
\cite{vatansever1, vatansever2}.  After a complete detailed
analysis, the studied system shows that  the first order phase
transitions as well as dynamic tricritical points (DTCPs)
disappear while the reentrant phenomena exists depending on the
value of  quenched bond dilution parameters $p$
\cite{vatansever1}. It is also emphasized that
frequency dispersions of hysteresis  loop area, remanence and coercivity
strongly depend on the  quenched bond randomness,  as well as applied
field amplitude and  oscillation frequency \cite{vatansever2}.
Furthermore, it is beneficial to emphasize that
Monte Carlo simulation studies concerning the nonequilibrium
behaviors and universality aspects of different type Ising spin
glasses systems take part in literature \cite{anderson, katzgraber,
nakamura, roma}. Even though there exists a limited number of
studies including the disorder effects under the  time dependent
oscillating magnetic field, the dynamic nature  of the $\pm J$
Ising model which would exhibit an exotic and unusual dynamic
behaviors has not yet been investigated. From this point of view,
in this work, we intend to probe the effects of the random bond
dilution process on the kinetic Ising model in the presence
of a time-dependent oscillating external magnetic field by using
the effective field theory with correlations based on the exact
Van der Waerden identity for a spin-1/2 system.

The remainder of the study is as follows:  The dynamic equation of motion
and dynamic order parameter (DOP) of the  kinetic  $\pm J$ Ising model
are described in  Sect. \ref{formulation}. The numerical results and
related  discussions are  given in Sect. \ref{discussion},
and finally Sect. \ref{conclusion}  contains our conclusions.
\section{Formulation}\label{formulation}
We consider a two dimensional $\pm J$ Ising model defined
on a square lattice, and the considered system is subjected to a
periodically oscillating magnetic field.
The time dependent Hamiltonian describing our model
of magnetic system can be written as
\begin{equation}\label{eq1}
\mathcal{H}=-\sum_{\langle ij \rangle} J_{ij}S_{i}S_{j}-
\sum_{i=1}^{N}H_{i}(t)S_{i}
\end{equation}
where $S_{i}=\pm1$ is the Ising spin variables and  first
sum in Eq. (1) is over the nearest neighbor pairs of spins. We assume
that the nearest neighbor interactions are randomly diluted on the lattice
according to the probability distribution function
\begin{equation}\label{eq2}
P(J_{ij})=p\delta(J_{ij}-J)+(1-p)\delta(J_{ij}+J).
\end{equation}
The time dependent magnetic field is as following
\begin{equation}\label{eq3}
H_{i}(t)=h\cos(\omega t)
\end{equation}
here, $h$ and $\omega$ are amplitude and angular frequency of the external
field, respectively.
In order to describe the dynamical evolution of the system,
we follow a Glauber-type stochastic process \cite{glauber} at a
rate of $1/\tau$ which represents the transitions per unit time, and
the dynamical equation  of motion can be obtained  by using the
master equation as follows:
\begin{equation}\label{eq4}
\tau\frac{d\langle S_{i}\rangle }{dt}=-\langle S_{i}\rangle+\left\langle
\tanh\left[\frac{E_{i}+H_{i}(t)}{k_{B}T}\right] \right\rangle
\end{equation}
where  $E_{i}=\sum_{j}J_{ij}S_{j}$ is the local field acting on the
lattice $i$, and $k_{B}$ and $T$ denote the Boltzmann constant and
temperature, respectively. If we apply the differential operator technique
\cite{honmura, kaneyoshi1} in Eq. (4) by taking into an account
the random configurational   averages we get
\begin{equation}\label{eq5}
\frac{dm}{dt}=-m+\left \langle \left \langle \prod_{j=1}^{q=4}
\left[A_{ij}+S_{j}B_{ij}\right]\right \rangle \right \rangle_{r} F(x)|_{x=0}
\end{equation}
here, where $A_{ij}=\cosh(J_{ij}\nabla)$, $B_{ij}=\sinh(J_{ij}\nabla)$
and $m=\langle \langle S_{i}\rangle \rangle$ indicates the average
magnetization. $\nabla=\frac{\partial}{\partial x}$ is the
differential operator, and the inner $\langle \cdots \rangle $
and the outer $\langle  \cdots \rangle_{r} $ brackets represent thermal
and configurational averages, respectively. When the right-hand side of
Eq. (\ref{eq5}) is expanded, the multispin correlation
functions appear. The simplest approximation,
and one of the most frequently adopted is to decouple
these correlations according to
\begin{equation}\label{eq6}
\langle \langle S_{i}S_{j} \cdots S_{l}\rangle \rangle_{r} \cong
\langle \langle S_{i} \rangle \rangle_{r}\langle \langle S_{j}
\rangle \rangle_{r} \cdots \langle \langle S_{l} \rangle \rangle_{r},
\end{equation}
for $i\neq j \cdots \neq l$ \cite{tamura}. By making use of Eq. (\ref{eq6}),
the right hand side of Eq. (\ref{eq5}) can be easily expanded and then,
the following dynamic effective field equation of motion for the magnetization
of the $\pm J$ Ising model can be found as
\begin{equation}\label{eq7}
\frac{dm}{dt}=-m+\sum_{i=0}^{q=4}\ell_{i}m^{i}
\end{equation}
where the coefficients $\ell_{i}$ can be readily
determined by employing the mathematical relation
$\exp(\alpha \nabla)F(x)=F(x+\alpha)$. Eq. (\ref{eq7})
is a kind of initial value problem, and generally,
the solutions act a stable function after a certain transition
regime. In addition to these, the studied magnetic system
exhibit two type solutions, and the first one is called
symmetric solution that obeys the following property:
\begin{equation}\label{eq8}
m(t)=-m(t+\pi/\omega).
\end{equation}
In this type of solution, the time dependent magnetization oscillates
around zero value which corresponds to dynamically paramagnetic phase (P).
The second one is asymmetric solution where the magnetization oscillates
around a nonzero value which indicates the existence a dynamically
ferromagnetic phase (F). According to our solutions,
the observed behavior of the magnetization is independent of
the choice of the initial value of magnetization. On the other hand,
depending on Hamiltonian parameters and also initial value of
time dependent magnetization, there exist coexistence
regions (F+P phases) in the phase diagrams in temperature
versus field amplitude plane.

The time average magnetization over a full cycle of the
external magnetic field acts as DOP
which is defined as follows \cite{tome}
\begin{equation}\label{eq9}
Q=\frac{\omega}{2\pi}\oint m(t)dt
\end{equation}
where $m(t)$ periodic and stable function.

\section{Results and Discussion}\label{discussion}
In this section, we will focus our attention on the DPT
properties of $\pm J $ Ising model under a time dependent magnetic
field source. Here, we discuss how the bond randomness affects the
dynamic evolution of studied system in the vicinity of dynamically
order and disorder transition temperature. On the other hand,
it is known fact that it is not possible to obtain the free energy
for kinetic models in the presence of time-dependent external fields.
Hence, in order to determine the type of DPT
(first or second order), it is convenient to check the temperature dependence
of DOP. Namely, if the DOP decreases continuously to zero in
the vicinity of critical temperature, this transition is classified
as second order, whereas if it vanishes discontinuously, then
the transition is assumed to be first order. Keeping in this mind, and after detailed
analysis with a great success to construct dynamic phase boundary (DPB) that
separates the dynamic  ordered phases from the dynamic disordered phases,
it is shown that the behavior of the system changes dramatically with varying
Hamiltonian parameters. In order to investigate these  effects on
the dynamic behavior of the considered system, we
plot the DPB in various  planes.

\begin{figure}[!here]
\begin{center}
\includegraphics[width=8.0cm,height=7.5cm]{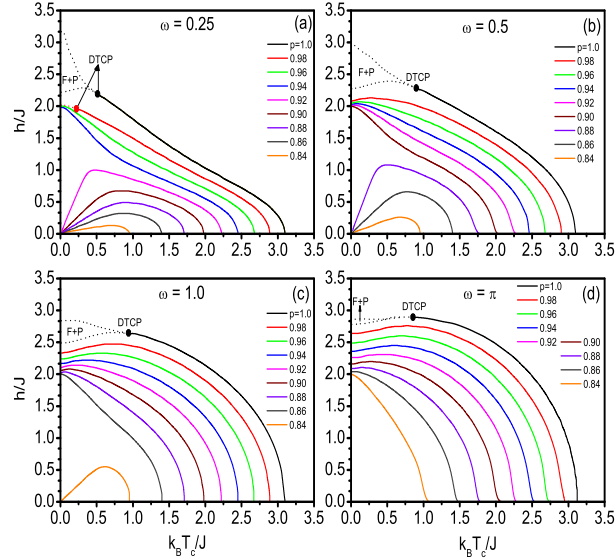}
\caption{Phase diagrams of the kinetic $\mathrm{\pm J}$ Ising model in
the $(\mathrm{k_{B}T_{c}/J-h/J})$ plane for selected bond
randomness values with  (a) $\mathrm{\omega=0.25}$,
(b) $\mathrm{\omega=0.5}$, (c) $\mathrm{\omega=1.0}$
and (d) $\mathrm{\omega=\pi}$. Solid (dotted) lines correspond
to second- (first-) order DPTs,
and $\bullet$ symbols represent DTCPs. }\label{fig1}
\end{center}
\end{figure}

Fig. \ref{fig1} (a-d) represents the DPT lines in reduced
temperature and applied field amplitude plane $(k_{B}T_{c}/J-h/J)$  for varying $p$
values and selected  applied field  frequencies such as $\omega=0.25 (a), 0.5 (b),
1.0 (c)$ and $\pi (d)$. Here, the full and dotted lines indicate the
continuous and discontinuous phase transition points, respectively, and
also the solid circle shows the DTCP. It is observed that a
reentrant behavior of second order where the two  successive  second order phase
transitions appears in the relatively low applied field amplitude and low
temperature regions for low $p$ values.
\noindent Another important result of our study
is that an increase in the value of $\omega$ leads to annihilate the
reentrant behavior and also to expand the dynamically ferromagnetic region.
On the other hand, with the increasing value of  $p$ the dynamic
evolution of the $\pm J$ Ising system begins to resemble the pure
kinetic Ising model, and a dynamically first order reentrant
behavior with a coexistence region (F+P) phase appears
where the critical properties of the system depends on the initial
value of the magnetization. Namely, if initial value of the $m(t)$ is selected
to be zero, the time dependent magnetization oscillates around zero value
which corresponds to P whereas
if it is considered to be nonzero, in this time,  $m(t)$
oscillates around a nonzero value, and this type of behavior corresponds to
F. It is possible to make an inference
that the  increasing field frequency causes a growing phase delay between
the magnetization  and magnetic field and this makes the occurrence
of the DPT difficult,  as a result of this  mechanism
the DPB gets wider and  also the reentrant  behavior
disappears at the low  temperature and low amplitude regions.
\begin{figure}[!here]
\begin{center}
\includegraphics[width=8.0cm,height=7.5cm]{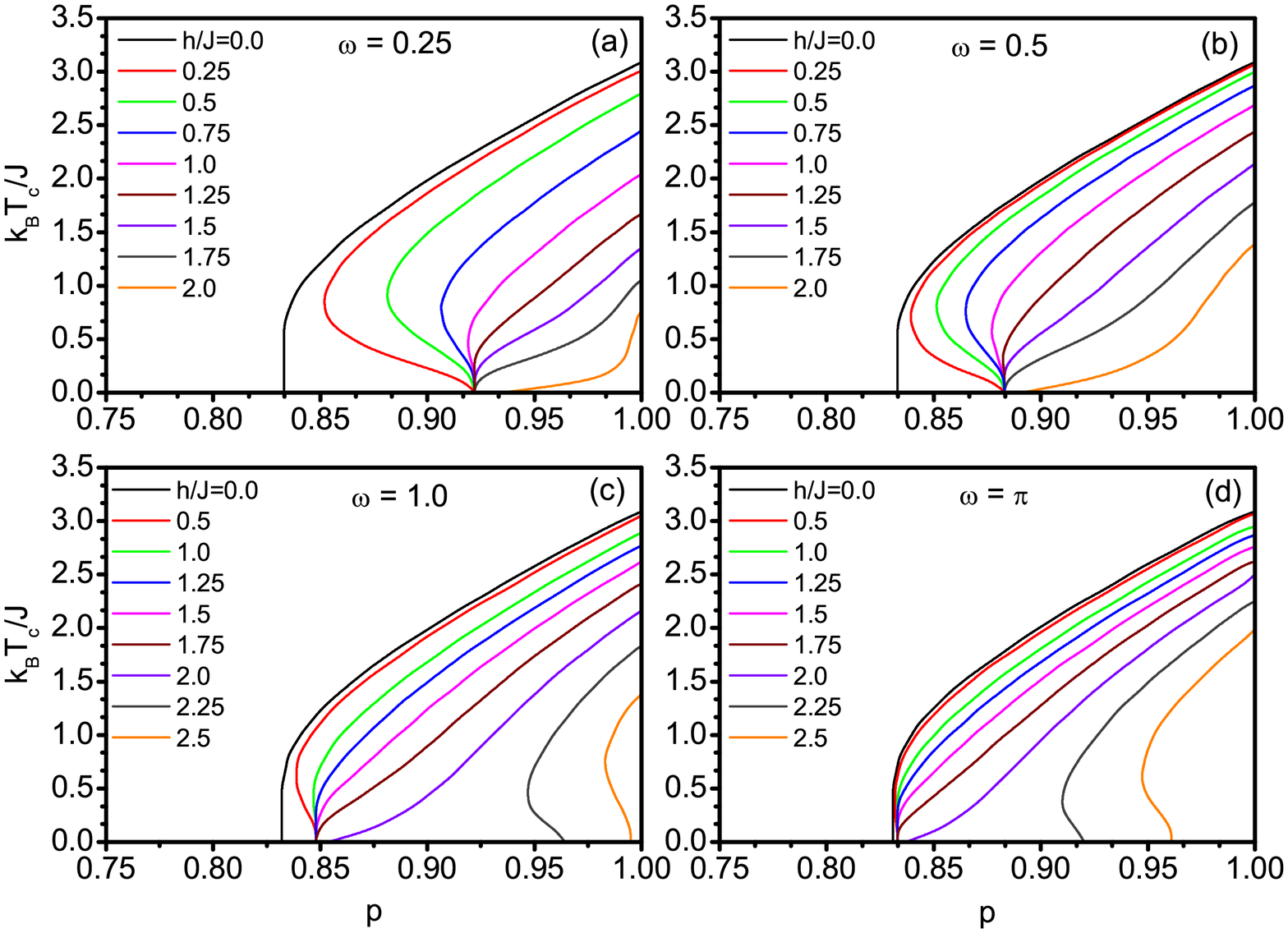}
\caption{Phase diagrams of the kinetic $\mathrm{\pm J}$ Ising model in
the $(\mathrm{p-k_{B}T_{c}/J})$ plane for varying reduced magnetic field
values with (a) $\mathrm{\omega=0.25}$, (b) $\mathrm{\omega=0.5}$,
(c) $\mathrm{\omega=1.0}$ and  (d) $\mathrm{\omega=\pi}$.}\label{fig2}
\end{center}
\end{figure}

Fig. \ref{fig2} (a-d) illustrates the DPT
lines in  $p$ and reduced temperature  plane $(p-k_{B}T_{c}/J)$
for varying $h/J$ values and selected applied field
frequencies such as $\omega=0.25 (a), 0.5 (b), 1.0 (c)$ and $\pi (d)$.
One can easily see that as $p$ value decreases then the ferromagnetic
phase region  gets narrower. According to our model,
as the concentration of active ferromagnetic bonds
is decreased from one then the antiferromagnetic  exchange interactions
appear, and due to the competition between ferromagnetic
and antiferromagnetic exchange interactions, energy contribution
which comes from spin-spin interactions gets smaller. Hence, the system
can undergo a DPT at lower critical temperatures,
and also  the dynamically ferromagnetic regions get narrower. On the other
hand, the case of  $h/J=0.0$  corresponds to static $\pm J$ Ising model,
in which there is no energy contribution originating from
magnetic field. Our investigations calculated for this applied field amplitude
indicates that the critical  percolation value is $p_{c}=5/6$. At
the critical percolation value the concentration of antiferromagnetic
bonds equal to $1/6\cong 0.167$, and it is worth noting that this value is in good
agreement with previously published works \cite{almeida, fittipaldi,
 domany, kirkpatrick, sarmento, katsura}. Another important finding is that
increasing $h/J$ values reduce the dynamic critical
temperature whereas as $\omega$ increases then the F
region gets wider in the related plane.

\begin{figure}[!here]
\begin{center}
\includegraphics[width=8.0cm,height=7.5cm]{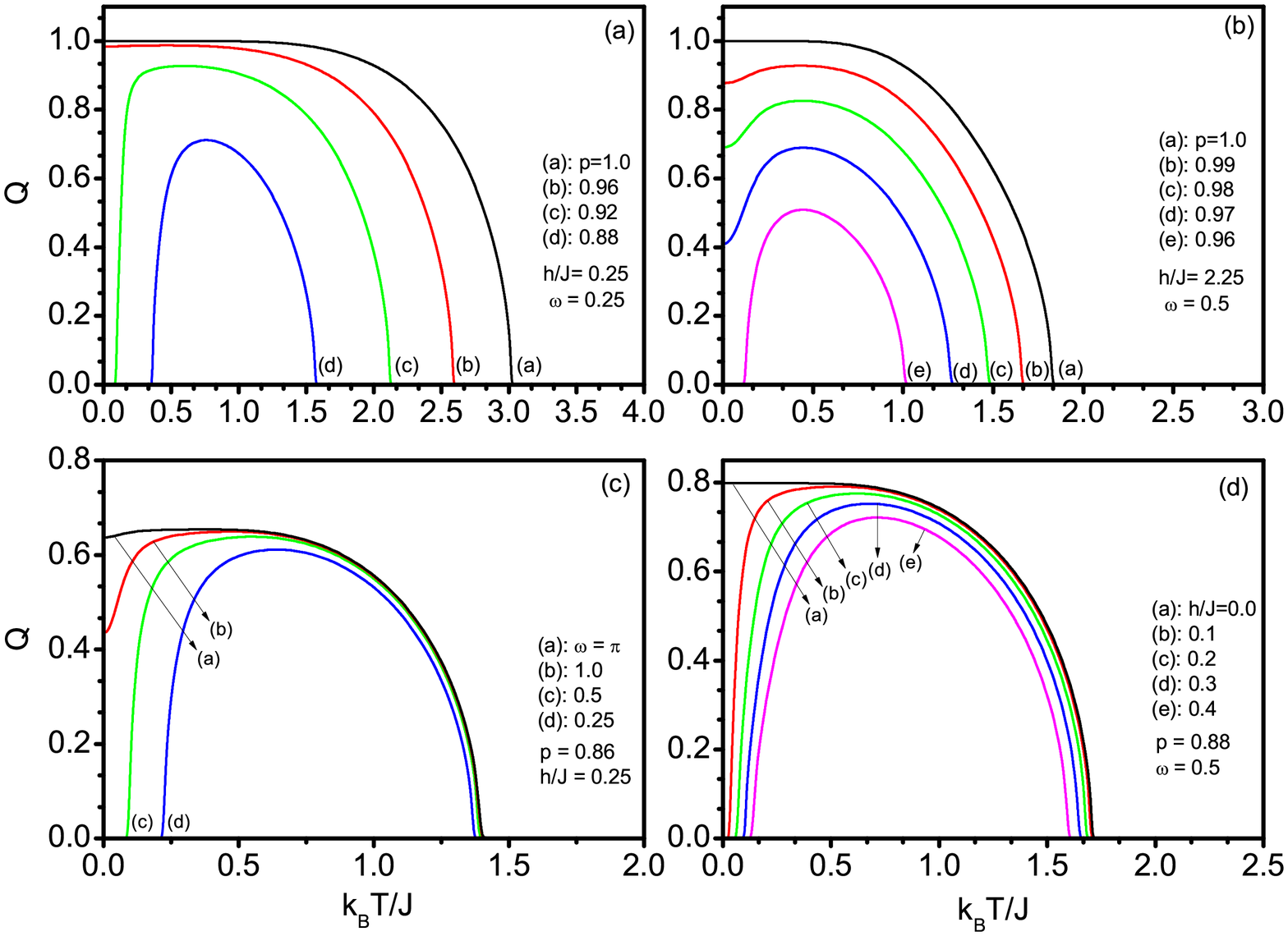}
\caption{Thermal variations of the DOP
curves as functions of the selected Hamiltonian parameters
$\mathrm{h/J}$, $\mathrm{\omega}$ and $\mathrm{p}$ where the
letters on the curves  denote the value of the
corresponding Hamiltonian  parameter.}\label{fig3}
\end{center}
\end{figure}

In Figs. \ref{fig3}(a-d), in order to elucidate the effects
of  the quenched random bond process on the studied system
in detail,  we give the thermal variations of DOP
corresponding to DPBs depicted
in Fig. \ref{fig1} and Fig. \ref{fig2} for a combination of
Hamiltonian  parameters. Based on the upper left and right
panels in Fig. \ref{fig3}, it can be easily said that as
the active ferromagnetic bond concentration decreases then
the system tends to exhibit reentrant behavior of second
order. Here, two successive DPTs exist,
and the first one occurs from a disordered phase to an
ordered phase at relatively low temperature regions,
while the other one takes place from an ordered phase to a
disordered phase at higher temperature regions
for decreasing $p$ values. It is also
possible to mention that the aforementioned
situation is valid for both weak and strong
frequency (or amplitude) values which can
be clearly seen from the upper left and right
panels in Fig. \ref{fig3}. One of
the most interesting findings is that the competition
between ferromagnetic and antiferromagnetic
bonds gives rise to the existence of
different type magnetization profiles. According to
N\'{e}el nomenclature \cite{neel, strecka}, it is possible to classify
the thermal variation of the magnetization curves in certain categories.
Based on this classification scheme, one can see
from Fig. \ref{fig3}(b) that considered system
exhibits Q-type, L-type  and P-type  magnetization
behaviors depending on Hamiltonian parameters.
Furthermore,  the variations of the DOP with temperature curves
as a function of the applied field frequency are shown
in the lower left panel in Fig. \ref{fig3} for $p = 0.86$
and $h/J = 0.25$. As shown in this figure, value of the DOP
 decreases when the frequency approaches
the static case. The physical backgrounds underlying the
behaviors found in Fig. \ref{fig3}(c) are identical to
those emphasized in above discussions. Therefore, we will
not discuss these interpretations here.  In addition to these,
the thermal variations of the DOP at
various $h/J$ values show  that depending on the applied field
amplitudes the system exhibits second order reentrant phenomena
in Fig. \ref{fig3}(d).

\begin{figure}[!here]
\begin{center}
\includegraphics[width=8.0cm,height=6.5cm]{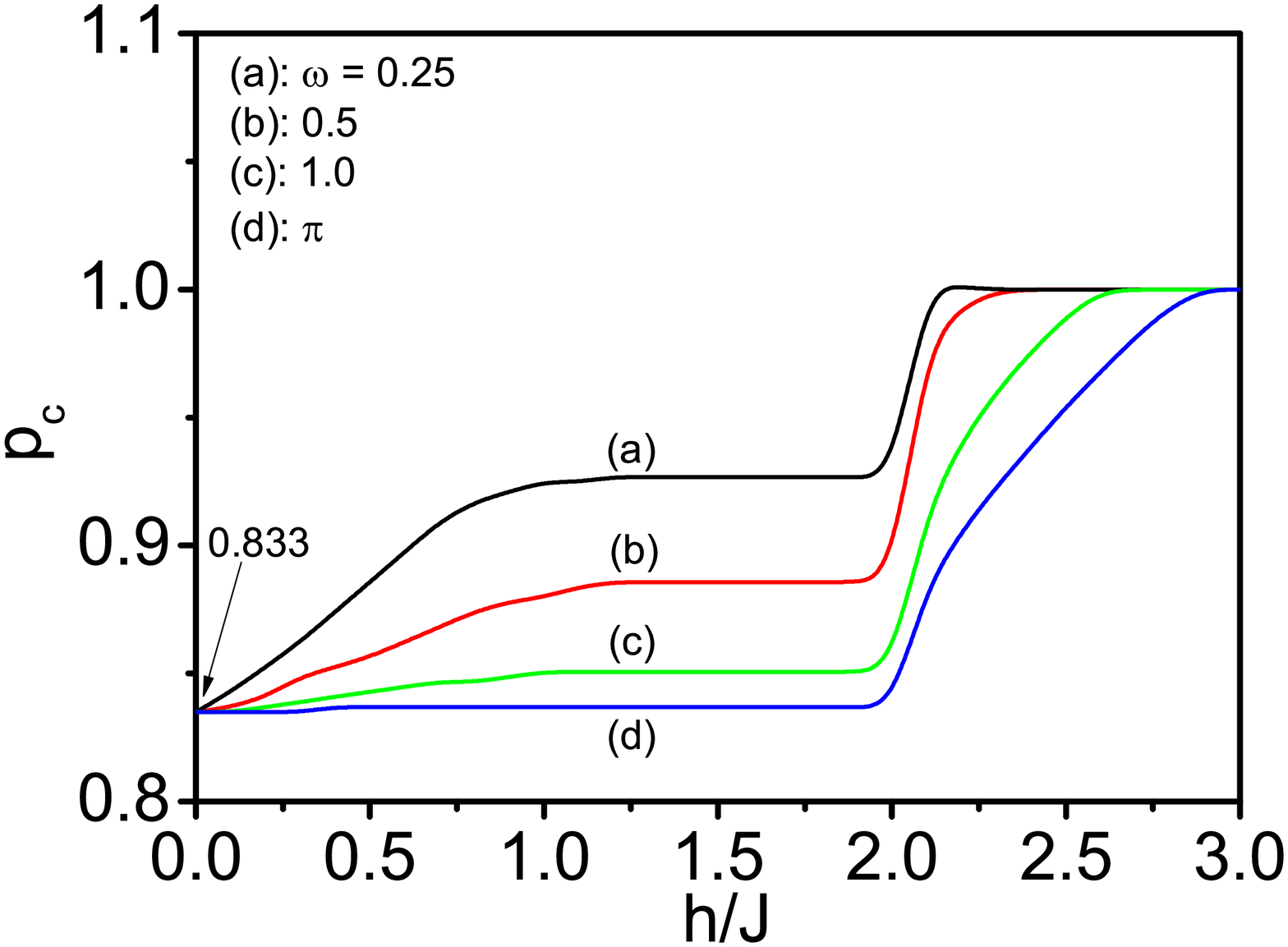}
\caption{Variation of the bond
percolation threshold $\mathrm{p_{c}}$ as a
function of the external magnetic field
amplitude $\mathrm{h/J}$ for some selected
values of the oscillation frequency $\mathrm{\omega}$.
The letters on each curve denote the
frequency value of the external field}\label{fig4}
\end{center}
\end{figure}

In the following analysis, let us investigate the variation
of the bond percolation threshold $p_{c}$ as a function
of applied field amplitude $h/J$ with some selected
values of $\omega$ which is depicted in Fig. \ref{fig4}.
We found that at relatively small oscillation frequency
values such as $\omega=0.25, 0.5$ and $1.0$, value
of $p_{c}$ increases gradually  then exhibits a plateau
which gets wider as $\omega$  increases. Another
important findings related to the percolation
investigation is that after a certain $h/J(=2.0)$ value,
$p_{c}$ value increases rapidly and
saturates at $p_{c}=1.0$. At this point, we should
indicate that for $p<p_{c}$, there is no
infinite cluster causing to ordered
phase and no  phase transition whereas there is
an infinite cluster for $p>p_{c}$ and critical
temperature rises from zero. On the other hand,
it is possible to say that the critical percolation
value $p_{c}$ strongly depends on a kind of
competition effect which originates
from the collaboration of the ferromagnetic and
antiferromagnetic  bond interactions with the applied
field frequency against the  amplitude of the
external field. Moreover, as we mentioned above, the
case of  $h/J=0.0$  corresponds to static $\pm J$ Ising
model, and the numerical calculations show that
percolation value is $p_{c}=5/6(\cong0.833)$ which is
independent of $\omega$.

\begin{figure}[!here]
\begin{center}
\includegraphics[width=8.0cm,height=6.5cm]{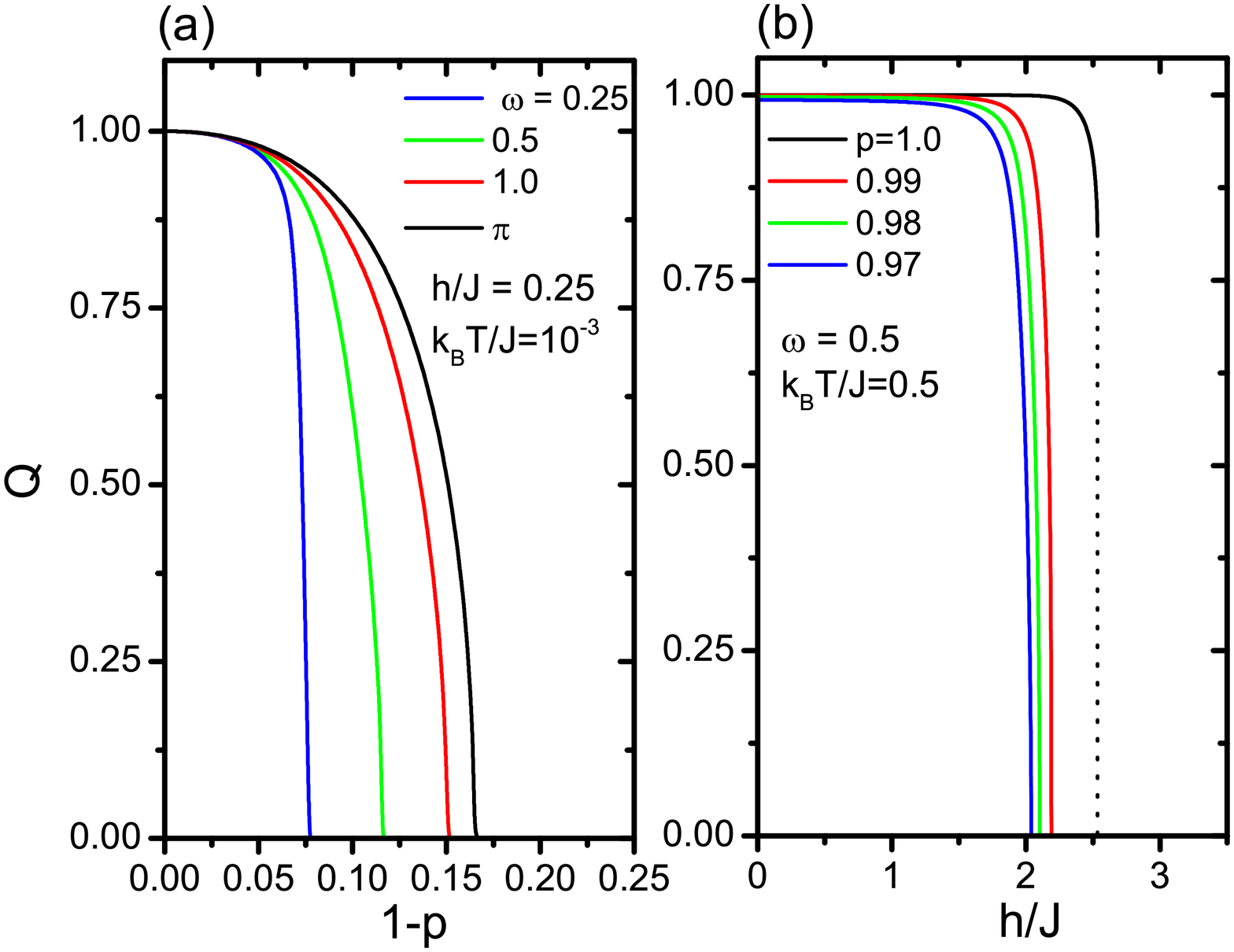}
\caption{Bond randomness dependencies of the DOP
 for fixed temperature and applied field
amplitude at various frequencies of field (a).
Magnetic field variation of the DOP for
considered Hamiltonian parameters (b).}\label{fig5}
\end{center}
\end{figure}

The bond randomness and external field amplitude
dependence of the DOP are shown
in Fig. \ref{fig5} (a-b) for a considered
combination of Hamiltonian parameters. It is obvious
from the  Fig. \ref{fig5} (a) that an increment on
the active  antiferromagnetic bond concentrations
$(1-p)$ tends to destruct the dynamically
ordered phase for value of $k_{B}T/J=10^{-3}$ and
$h/J=0.25$. We notice that the aforementioned
situation is dependent on applied field frequency,
and with the decreasing value of $\omega$, the
dynamically order-disorder transition point
shifts to lower value of active
antiferromagnetic bond concentrations.
At this point, we can also mention that there exists
a good agreement in a formal manner between
our Fig. \ref{fig5}(a) and Fig. 5 for $\alpha=-1$ of
Ref. \cite{sarmento} where static properties of a
quenched random bond Ising ferromagnet with anisotropic
coupling constants are discussed.  Moreover,
in Fig. \ref{fig5} (b), it is observed that the
competition between the  ferromagnetic and
antiferromagnetic exchange  interactions give rise to
destruct the dynamically first order phase transition
as well as to disappear the DTCP
at various $p$  values and  for fixed reduced temperature
and  field frequency  such as $k_{B}T/J=0.5$ and $0.5$,
respectively.

\begin{figure}[!here]
\begin{center}
\includegraphics[width=8.0cm,height=6.5cm]{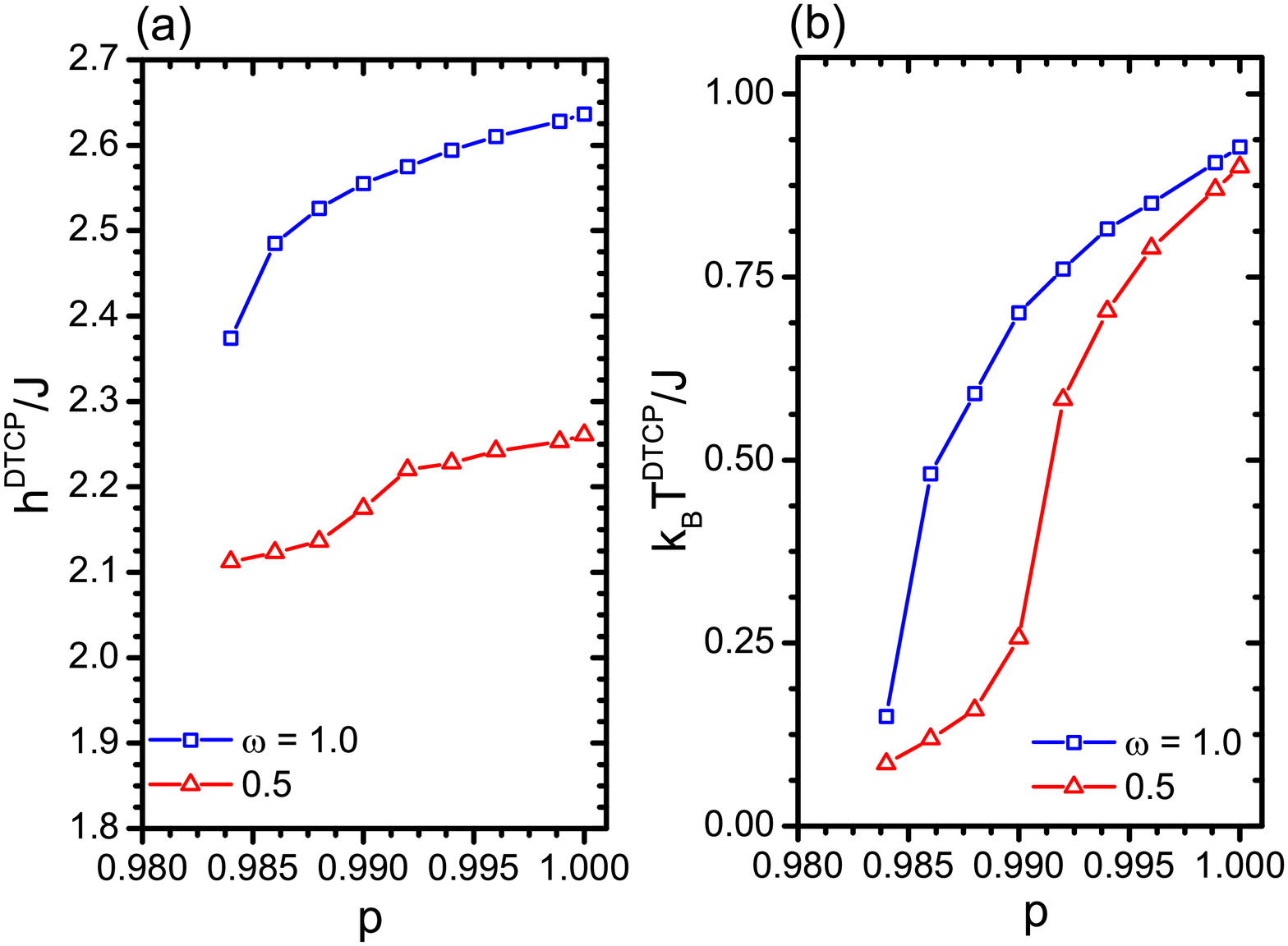}
\caption{Variations of DTCP coordinates
(a) $\mathrm{h^{DTCP}/J}$ and (b) $\mathrm{k_{B}T^{DTCP}/J}$ as a
function of bond randomness $\mathrm{(p)}$ for considered values of
oscillating field frequencies such as  $\mathrm{\omega}=0.5$ and $1.0$.}\label{fig6}
\end{center}
\end{figure}

As a final investigation, in Figs. \ref{fig6}(a) and
\ref{fig6}(b),  variations of DTCP coordinates
$\mathrm{h^{DTCP}/J}$ and $\mathrm{k_{B}T^{DTCP}/J}$
with respect to  the bond randomness are plotted for selected
oscillating field frequencies such as $\omega=0.5$ and $1.0$, respectively.
As seen in these figures, both the $\mathrm{h^{DTCP}/J}$ and
$\mathrm{k_{B}T^{DTCP}/J}$ diminish with decreasing active
ferromagnetic bond concentration, and the DTCP disappears
after a certain value. It is also possible to notice
that the coexistence region in the phase
diagrams gets narrower and  disappear with decreasing
$p$ values. Besides, the coordinates of DTCPs
explicitly depend on the applied field frequency,
and one can easily deduce from the figures that the
ordered phase regions and also positions
of the DTCPs in related planes tend to
expand because increasing field
frequency gives rise to a growing
phase delay between the magnetization and
magnetic field and this makes  the occurrence of
the DPT difficult.

\section{Concluding Remarks}\label{conclusion}
In conclusion, we have looked for an answer how the time dependent
oscillating magnetic field affects the DPT
properties of  $\pm J$ Ising model within the framework of effective
field theory on a two dimensional square lattice. In our studied
model the spin-spin exchange
interaction has a probability $p$ and $1-p$ of taking on values $+J$ and $-J$,
respectively. For this purpose, the Glauber type stochastic process has been
used  with a great success to determine the time  evolution of
competing the magnetic system. After a detailed  analytical and
numerical operations, the DPB separating the dynamically
ordered and disordered phases has been obtained in different planes
by benefiting from thermal variations of DOPs at
various Hamiltonian parameters. It is found  that the competition
between $\pm J$ interactions causes the reentrant phenomena for
some certain values of amplitude and frequency of the
external applied field, and the ferromagnetic
phase regions get expanded with decreasing amplitude which
is more evident at low frequencies. One of the
most important findings is that the bond randomness leads to
destruct the dynamically first order phase transition as well as
to disappear the coexistence regions.  Namely, after a certain
value of ferromagnetic or antiferromagnetic concentration,
the first order phase transitions turn  into the second
order phase transitions and consequently, DTCPs
disappear for all frequency values. Furthermore, it is observed
according to N\'{e}el nomenclature \cite{neel, strecka} that,
the magnetization  curves of kinetic $\pm J$ Ising model have
been found to obey Q-type, L-type  and P-type classification
schemes  under certain conditions (see Fig. \ref{fig3}(b)).
We should also emphasize that
the DPBs in related planes have been constructed
by making use of the temperature variations of DOP,
and  our calculated phase boundaries
do not include spin glass phase
because there is no infinite cluster
causing to ordered phase for $p<p_{c}$,
in this region the DOPs only indicate
the existence of dynamically
disordered phases. We think that focusing on
hysteresis and loop area properties of the
studied system may be good and powerful ways to determine and
investigate  of the spin glass
behavior. Recently, this  type of
calculation has been done and spin glass
phase properties of the $d=3$ $\pm J$ Ising
model have been elucidated in detail
within the framework of hard spin
mean field theory \cite{sariyer}.

On the other hand, it is a well known fact that effective
field theory takes the standard mean field predictions one
step forward by taking into account the single-site
correlations which means that the thermal fluctuations
are partially considered  within the framework of effective field theory.
It is possible to mention that the effective field theory
can be  successfully applied to such non-equilibrium systems
in the presence of bond randomness, however, the true
nature of the physical facts underlying the origin of the
coexistence phase and also dynamic first order transitions
may be further understood by benefiting from a well
defined powerful method such as Monte Carlo simulation technique.
\section*{Acknowledgements}
The numerical calculations reported in this paper were
performed at T\"{U}B\.{I}TAK ULAKBIM (Turkish agency),
High Performance and Grid Computing Center (TRUBA
Resources).

\end{document}